\documentclass{elsart}
\usepackage{amssymb,amsfonts}
\usepackage{graphicx}
\newcommand{\trace}{\mathop{\rm Tr}\nolimits}

\newcommand{\cT}{{\mathcal T}} 
\newcommand{\cR}{{\mathcal R}} 

\newcommand{\id}{\mathbb{I}}

\newcommand{\be}{\begin{equation}}
\newcommand{\ee}{\end{equation}}
\newcommand{\bea}{\begin{eqnarray}}
\newcommand{\eea}{\end{eqnarray}}
\newcommand{\beas}{\begin{eqnarray*}}
\newcommand{\eeas}{\end{eqnarray*}}

\newtheorem{theorem}{Theorem}

\newtheorem{corollary}{Corollary}

\newtheorem{proposition}{Proposition}

\newcount\minute
\newcount\hour
\def\currenttime{%
    \minute\time
    \hour\minute
    \divide\hour60
    \the\hour:\multiply\hour60\advance\minute-\hour\the\minute}
\begin{document}
\begin{frontmatter}
\title{On the asymmetry of the relative entropy}
\author{Koenraad M.R.\ Audenaert}
\address{
Department of Mathematics,
Royal Holloway, University of London, \\
Egham TW20 0EX, United Kingdom}
\ead{koenraad.audenaert@rhul.ac.uk}
\date{\today, \currenttime}
\begin{abstract}
The quantum relative entropy $S(\rho||\sigma)$ is a widely used dissimilarity measure between quantum states, but it has the
peculiarity of being asymmetric in its arguments.
We quantify the amount of asymmetry by providing a sharp upper bound in terms of two parameters:
the trace norm distance between the two states, and the smallest of
the smallest eigenvalues of both states. 
The bound is essentially the asymmetry between two binary distributions governed by these two parameters.
\end{abstract}

\end{frontmatter}
\section{Introduction\label{sec:intro}}
The quantum relative entropy between two quantum states $\rho$ and $\sigma$,
$S(\rho||\sigma)=\trace\rho(\log\rho-\log\sigma)$, is a non-commutative generalisation
of the Kullback-Leibler divergence (KLD) $D_{KL}(p||q)$ between probability distributions $p$ and $q$. 

Just as the KLD, the relative entropy is not a true metric distance, first and foremost because it is not symmetric in its arguments.
In essence, this asymmetry is not a deficiency but a feature, arising from the inherent asymmetry in the mathematical models from which
both concepts emerge.
For example, $D_{KL}(p||q)$ can be interpreted as the number of extra bits required to encode a bitstream assuming it comes from a source 
with distribution $q$, where in fact the source is governed by distribution $p$.
In the setting of hypothesis testing, where under hypothesis $H_0$ a random variable is distributed according to $p$
and under hypothesis $H_1$ according to $q$,
$D_{KL}(p||q)$ can be interpreted as the expected `weight of evidence' per sample in favour of $H_1$ and against $H_0$.
To clarify the asymmetry here, we simply quote F.~Bavaud, who wrote, paraphrasing Popper \cite{bavaud09}: 
\begin{quote}
``The theory `All crows are black' is refuted by the single observation of a white crow, while the
theory `Some crows are black' is not refuted by the observation of a thousand white crows.''
\end{quote}

Be this as it may, the quantum relative entropy
is widely used as a quantitative measure of the dissimilarity between two quantum states \cite{ohya_petz}, not in the least because of its 
simplicity, its clear information theoretical meaning, and its nice mathematical properties.
In these applications, the asymmetry is just considered part of the price to be paid.

If one does not wish to pay this price, one way out is to replace the relative entropy by a symmetrisation \cite{nielsen}.
The symmetrised KL divergence is known as the Jeffreys divergence, or J-divergence:
$$
J(p,q) = D_{KL}(p||q) + D_{KL}(q||p).
$$
Likewise, one can define a quantum J-divergence as
$$
J(\rho,\sigma) = S(\rho||\sigma) + S(\sigma||\rho).
$$

The question addressed in this paper is: how much can the quantum J-divergence differ from the quantum relative entropy?
Or, asked differently, how great can the asymmetry in the quantum relative entropy be?
It is well-known that in the infinitesimal limit, for distributions that are infinitesimally close, the KLD becomes a true metric,
its Hessian being known as the Fisher information metric, and the same can be said about the quantum relative entropy.
For states that are sufficiently close, we can therefore expect the asymmetry to be small.

\begin{figure}
\includegraphics[width=6.8cm]{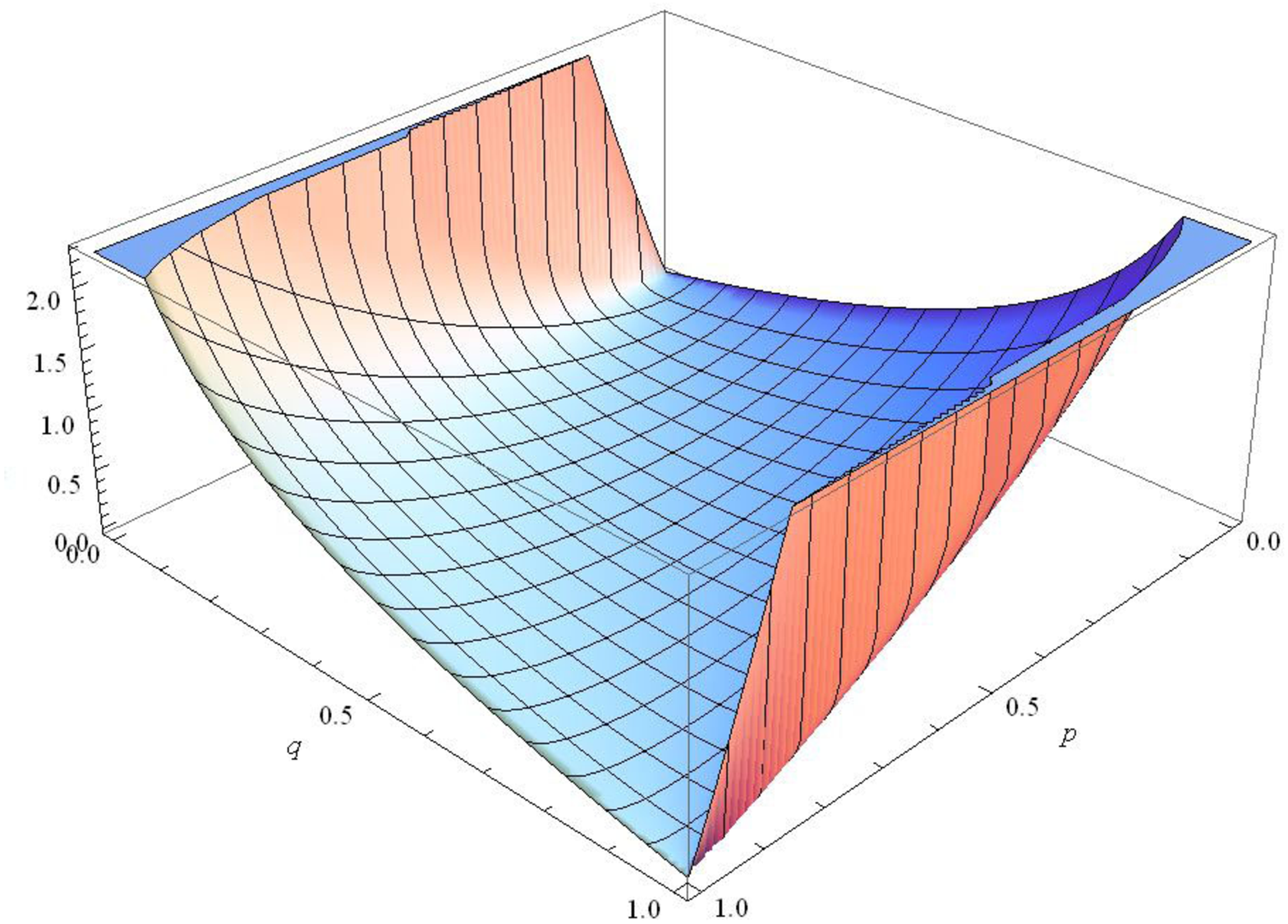}
\includegraphics[width=6.8cm]{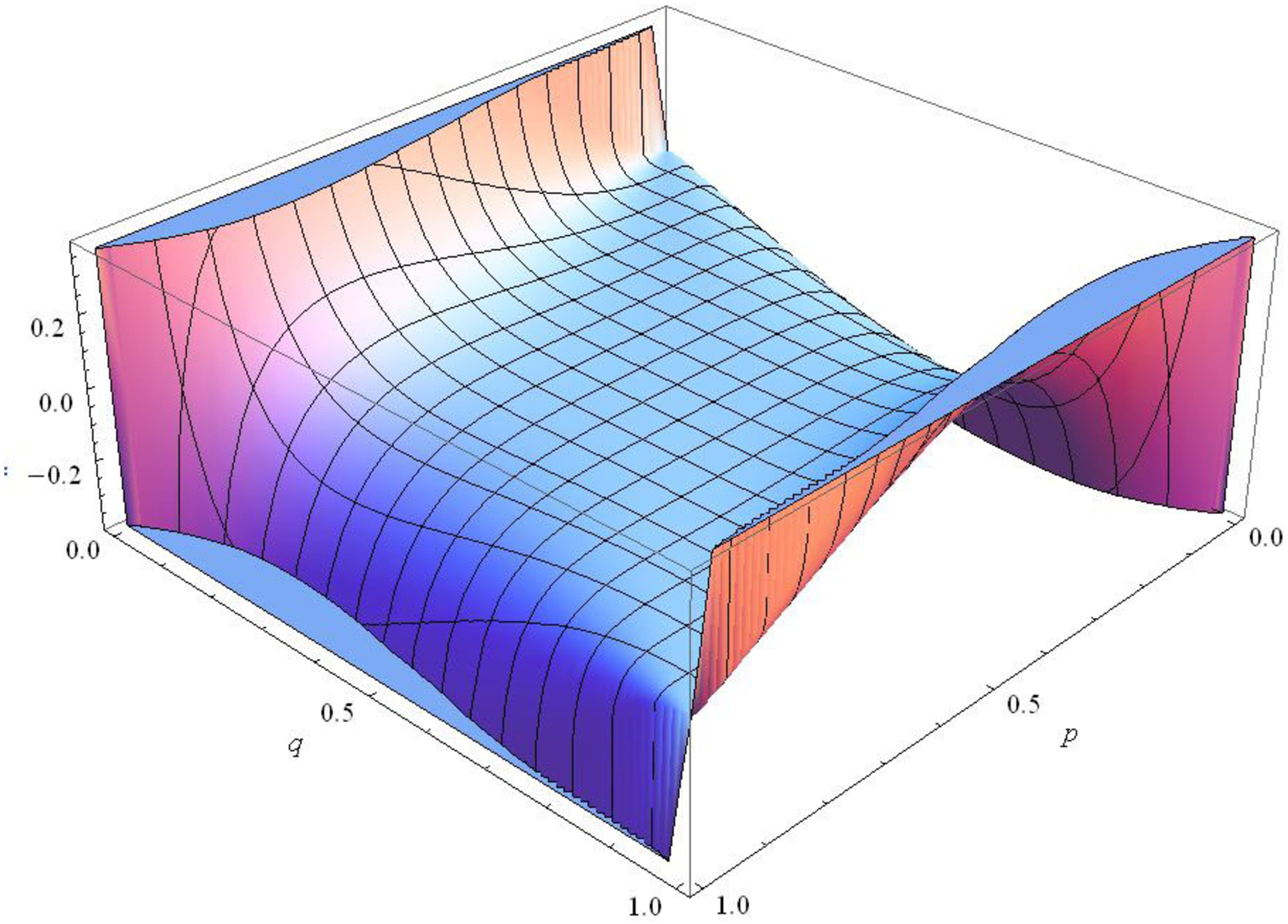}
\caption{
Kullback-Leibler distance between two binary distributions $(p,1-p)$ and $(q,1-q)$, and its asymmetry.
\label{fig:S}
}
\end{figure}

To make this statement more precise, we will first look at the simplest example of two binary distributions, $(p,1-p)$ and $(q,1-q)$.
Let us denote the KLD between these two binary distributions by the function 
$$
s_2(p||q) := p\log(p/q) + (1-p)\log((1-p)/(1-q)).
$$ 
A graph of this function is shown in  Fig.~\ref{fig:S}, along with a graph of its asymmetry
$s_2(q||p)-s_2(p||q)$.
The statement about the smallness of the asymmetry is partially 
corroborated by the presence of a relatively flat `plateau' in the middle of the graph. However, one also notices 
that for very small values of 
$p$ or $1-p$, the values of $p$ and $q$ have to be much closer together to keep the asymmetry small.

\begin{figure}
\includegraphics[width=12cm]{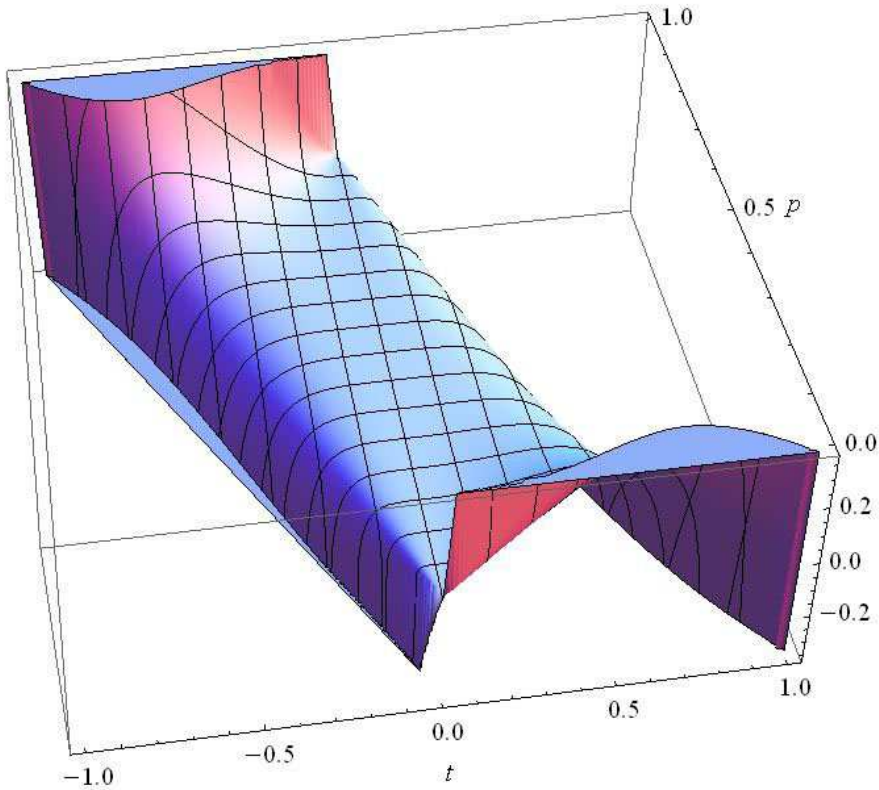}
\caption{
Graph of the asymmetry function $a(p,t)$.
\label{fig:axt}
}
\end{figure}

The asymmetry can be expressed in terms of the difference $t:=q-p$ by the function
\bea
a(p,t) &:=& s_2(p+t||p) - s_2(p||p+t) \\
&=& (2p+t)\log\left(1+\frac{t}{p}\right) + (2(1-p)-t)\log\left(1-\frac{t}{1-p}\right),
\eea
which is defined for $-1\le t\le 1$ and $\max(0,-t)\le p\le \min(1,1-t)$ (see Fig.\ \ref{fig:axt}).
A more qualitative statement about the flatness of this function can be made by considering
the Taylor series expansion of $a(p,t)$ as a function of $t$,
$$
a(p,t) = \left(p^{-2}-(1-p)^{-2}\right)\frac{t^3}{6} - \left(p^{-3}-(1-p)^{-3}\right)\frac{t^4}{6} +O(t^5).
$$
and noticing that the leading term is of order 3 in $t$.

The main technical contribution of the present paper is that the situation just considered for binary distributions is 
essentially universal and holds for distributions and quantum states of any (finite) dimension, provided 
the parameter $p$
in the asymmetry function $a(p,t)$ is replaced by  
the smallest of the smallest eigenvalues of $\rho$ and $\sigma$, and
the parameter $t$ is replaced by the trace norm distance $T$ between $\rho$ and $\sigma$.
Then the absolute value of the asymmetry function $|a(z,T)|$ is a sharp upper bound on the asymmetry
$A(\rho||\sigma) := |S(\sigma||\rho)-S(\rho||\sigma)|$.
\section{Main Results}
The bounds we prove here can be conveniently expressed using the function $a(p,t)$ defined in the Introduction.

\begin{theorem}\label{th:main}
Let $\rho$ and $\sigma$ be two density matrices, with trace distance $T=||\rho-\sigma||_1 /2$ and $\lambda_{\min}(\sigma)=x$.
Then
\be
S(\rho||\sigma) - S(\sigma||\rho) \le a(x,T).
\ee
\end{theorem}
The highly technical proof of this theorem is postponed to the last section.

\begin{corollary}
Let $\rho$ and $\sigma$ be two density matrices, with $T=||\rho-\sigma||_1 /2$,
$\lambda_{\min}(\sigma)=x$, and $\lambda_{\min}(\rho)=y$.
Then
\be
|S(\rho||\sigma) - S(\sigma||\rho)| \le a(\min(x,y),T).
\ee
\end{corollary}
\textit{Proof.}
As $|x|=\max(x,-x)$, an upper bound on $A(\rho,\sigma):=|S(\rho||\sigma) - S(\sigma||\rho)|$ is given by the pointwise maximum
of the bounds 
$A(\rho,\sigma) \le a(x,T)$ and
$A(\sigma,\rho) \le a(y,T)$,
where the latter is obtained from the former
by swapping the roles of $\rho$ and $\sigma$.
The statement follows from the facts that $A(\rho,\sigma)$ is symmetric in its arguments and that $a(x,T)$ 
is strictly decreasing in $x$ for any fixed $T\in[0,1]$.
\qed

Note that for $d$-dimensional
states $\rho$ and $\sigma$ under the restriction $\rho,\sigma\ge z$, their trace distance is bounded above by
$1-d z$.
\section{Proof}
\subsection{Preliminaries\label{sec:pre}}
The positive part of a self-adjoint operator $X$ is $X_+ := (X+|X|)/2$.
It features in an expression
for the trace norm distance between states:
\be
T(\rho,\sigma) := \frac{1}{2}||\rho-\sigma||_1 = \trace(\rho-\sigma)_+.
\ee

The following integral representation of the logarithm is well-known:
\be
\log x = \int_0^\infty ds \left(\frac{1}{1+s}-\frac{1}{x+s}\right),\quad x>0.\label{eq:intlog}
\ee
By functional calculus, this representation extends to positive operators $A\ge0$ as
\be
\log A = \int_0^\infty ds \left((1+s)^{-1}\id-(A+s)^{-1}\right).\label{eq:intlogA}
\ee
From this follows an integral representation for the derivative of the matrix logarithm: for $A\ge0$
$$
\cT_A(\Delta) := \frac{d}{dt}\Bigg|_{t=0} \log(A+t\Delta)
=\int_0^\infty ds\,\,(A+s\id)^{-1} \Delta (A+s\id)^{-1}.
$$
Just as the first derivative of the logarithm defines the linear operator $\cT$, we can also define
a quadratic operator $\cR$ via the second derivative \cite{lieb73}. For $A\ge0$ and $\Delta$ self-adjoint,
\bea
\cR_A(\Delta) &=& -\frac{d^2}{dt^2}\Bigg|_{t=0}\log(A+t\Delta) \\
&=& 2\int_0^\infty ds\,\,(A+s\id)^{-1}\Delta(A+s\id)^{-1}\Delta(A+s\id)^{-1}. \label{eq:intcR}
\eea
\subsection{A technical Proposition}
\begin{proposition}\label{prop:asym}
Let $\sigma$ be a finite dimensional density matrix with $x=\lambda_{\min}(\sigma)$.
Let $\Delta=\Delta_+-\Delta_-$ with $\trace\Delta_\pm=1$.
Let $t$ be a non-negative number such that $\sigma+t\Delta$ is also a density matrix.
Then
$$
\trace\Delta \cR_{\sigma+t\Delta}(\Delta) \le (x+t)^{-2}-(1-x-t)^{-2}.
$$
\end{proposition}
\textit{Proof.}
Denote $\rho=\sigma+t\Delta$.

The first step of the proof is a Fiedler-type argument\footnote{Named after Fiedler's technique 
used to prove a well-known result in matrix analysis,
see e.g.\ Th.~VI.7.1 in \cite{bhatia}.}
that $\trace\Delta \cR_\rho(\Delta)$ achieves
its maximal value when $\Delta$ and $\rho$ commute.
By the integral representation (\ref{eq:intcR}), we have
\beas
\trace\Delta \cR_\rho(\Delta)
&=& 2\int_0^\infty ds\;\trace(\Delta (\rho+s)^{-1} \Delta(\rho+s)^{-1} \Delta(\rho+s)^{-1}) \\
&=& 2\int_0^\infty ds\;\trace(\Delta (\rho+s)^{-1})^3.
\eeas
Now let $U$ be unitary and $\Delta=U\Delta_0 U^*$, and write $M=(\rho+s)^{-1}$. We first show that the maximum of
$\trace(\Delta (\rho+s)^{-1})^3$ over all unitary $U$ is obtained when $\Delta$ and $\rho$ commute.
Any unitary matrix $U$ can be written as $U=e^{tK}$, where $K$ is skew-Hermitian.
With this parameterisation,
\beas
\frac{d}{dt}\Big|_{t=0} \trace(\Delta M)^3
&=& \frac{d}{dt}\Big|_{t=0}\trace(e^{tK}\Delta_0 e^{-tK} M)^3\\
&=& 3\trace([K,\Delta_0]M\Delta_0 M\Delta_0 M) \\
&=& 3\trace(K((\Delta_0 M)^3-(M\Delta_0)^3)).
\eeas
Any extremal point of $\trace(\Delta M)^3$ is therefore characterised by the requirement that
$\trace(K((\Delta_0 M)^3-(M\Delta_0)^3))=0$
for all skew-Hermitian $K$. This amounts to the equation $(\Delta_0 M)^3=(M\Delta_0)^3$.
If we now make the assumption that $\Delta_0$ and $M$ are such that $\Delta_0 M$ has simple eigenvalues (which
is true for a dense subset),
this means that $\Delta_0 M=M\Delta_0$, too, i.e.\ $\Delta_0$ and $M$ must commute.

For such an extremal point to be a maximum, an additional condition must hold.
In a basis in which the eigenvalues of $M$ appear in decreasing order (in which both $M$ and $\Delta_0$
are diagonal) the diagonal elements of $\Delta_0$ must appear in decreasing order too; recall that
the eigenvalues of $M=(\rho+s)^{-1}$ are strictly positive for finite $s$.

Now this is so independently of the value of $s$. Therefore, the maximising $\Delta$ for the entire integral
$\int_0^\infty ds\;\trace(\Delta (\rho+s)^{-1})^3$ must commute with $\rho$, and in a basis
in which $\rho$ is diagonal and has its diagonal elements appearing in increasing order, the diagonal elements
of $\Delta$ must appear in decreasing order.

Now that the problem has been reduced to the commuting case,
we can simplify $\trace\Delta \cR_\rho(\Delta)$ to
$\trace \Delta^3\rho^{-2} = \trace \Delta_+^3(\sigma+t\Delta_+)^{-2} - \trace \Delta_-^3(\sigma-t\Delta_-)^{-2}$.
Given the conditions on $\Delta$, the ranks of its positive and negative parts must be between 1 and $d-1$.

Since $\lambda_{\min}(\sigma)=x$, we have $\sigma\ge x\ge x\Delta_+$.
Therefore,
$$
(\Delta_+^{-1}\sigma+t)^{-2}\le (x+t)^{-2},
$$
which immediately implies
\be
\trace \Delta_+^3(\sigma+t\Delta_+)^{-2} \le (x+t)^{-2}.\label{eq:asy1}
\ee

For any $A>0$ commuting with $\Delta_-$,
\beas
1 = \trace\Delta_- &=& \trace(\Delta_- A^{-2/3}A^{2/3}) \\
&\le& ||\Delta_- A^{-2/3}||_3\; ||A^{2/3}||_{3/2} \\
&=& (\trace\Delta_-^3 A^{-2})^{1/3}\;(\trace A)^{2/3},
\eeas
so that
$$
\trace\Delta_-^3 A^{-2} \ge (\trace A)^{-2}.
$$
Recall that the rank of $\Delta_-$ is at most $d-1$.
Applying this with $A$ equal to the restriction of $\rho=\sigma+t\Delta$ to the support of $\Delta_-$,
so that 
$$
\trace A\le \sum_{j=1}^{d-1} \lambda_j^\downarrow(\sigma-t\Delta_-)=1-x-t,
$$ 
we get
\be
\trace\Delta_-^3 (\sigma-t\Delta_-)^{-2} \ge (1-x-t)^{-2}.\label{eq:asy2}
\ee
Combining the two bounds (\ref{eq:asy1}) and (\ref{eq:asy2}), we get the bound of the proposition.
\qed

\subsection{Proof of Theorem \ref{th:main}.}
The proof of Theorem \ref{th:main} follows from the inequality of Proposition \ref{prop:asym} using three successive integrations.
Let $\Delta=(\rho-\sigma)/T$, so that $\Delta=\Delta_+-\Delta_-$ and $\Delta_\pm$ are density matrices.

Let us first perform the integration $\int_u^v dt$ on each side of the inequality
$$
\trace\Delta \cR_{\sigma+t\Delta}(\Delta) \le (x+t)^{-2}-(1-x-t)^{-2}.
$$
This gives
\beas
\lefteqn{\trace\Delta (\cT_{\sigma+u\Delta}(\Delta)-\cT_{\sigma+v\Delta}(\Delta))} \\
&\le& \int_u^v dt ((x+t)^{-2}-(1-x-t)^{-2}) \\
&=& (x+u)^{-1}-(x+v)^{-1}+(1-x-u)^{-1}-(1-x-v)^{-1}.
\eeas
Next, we perform the integration $\int_0^v du$:
\beas
\lefteqn{\trace\Delta (\log(\sigma+v\Delta)-\log\sigma -\cT_{\sigma+v\Delta}(v\Delta))} \\
&\le& \int_0^v du \left((x+u)^{-1}-(x+v)^{-1}+(1-x-u)^{-1}-(1-x-v)^{-1}\right) \\
&=& \log\frac{x+v}{v}-\frac{v}{x+v} +\log\frac{1-x}{1-x-v}-\frac{v}{1-v-x}.
\eeas
Now the left-hand side can be rewritten as
$$
\frac{d}{dv}\trace(2\sigma+v\Delta)(\log(\sigma+v\Delta)-\log\sigma).
$$
Performing the third and final integration $\int_0^T dv$ then yields
\beas
\lefteqn{\trace(2\sigma+T\Delta)(\log(\sigma+T\Delta)-\log\sigma)} \\
&\le& \int_0^T dv \left(\log\frac{x+v}{v}-\frac{v}{x+v} +\log\frac{1-x}{1-x-v}-\frac{v}{1-v-x}\right) \\
&=& (2x+T)\log(1+T/x) + (2(1-x)-T)\log(1-T/(1-x)).
\eeas
Noting that the left-hand side is just $S(\rho||\sigma) - S(\sigma||\rho)$ completes the proof.
\qed

\begin{ack}
This work was initiated by a question asked to the author by Jochen Rau.
\end{ack}


\begin{thebibliography}{9}
\bibitem{bavaud09} F.~Bavaud, ``Information Theory, Relative Entropy and Statistics'',
in: G.~Sommaruga (editor): \textit{Formal Theories of Information}. Lecture Notes in
Computer Science \textbf{5363}, Springer, Berlin, 54--78 (2009).
\bibitem{bhatia} R.~Bhatia, \textit{Matrix Analysis}, Springer.
\bibitem{lieb73} E.~Lieb, ``Convex Trace Functions and the Wigner-Yanase-Dyson Conjecture'',
Adv.\ Math.\ \textbf{11}, 267--288 (1973).
\bibitem{nielsen} F.~Nielsen, ``A family of statistical symmetric divergences
based on Jensen's inequality'', eprint arXiv 1009:4004 (2010).
\bibitem{ohya_petz} M.~Ohya and D.~Petz, \textit{Quantum entropy and its use}, Springer (1993).
\end{thebibliography}
\end{document}